\documentclass{article}
\usepackage{locata,amsmath,graphicx,url,times,textcomp,color}
\usepackage{multirow,stfloats}

\title{Evaluation of an open-source implementation of the SRP-PHAT algorithm within the 2018 LOCATA challenge}

\makeatletter
\def\name#1{\gdef\@name{#1\\}}
\makeatother
\name{\textit{Romain Lebarbenchon$^1$},
	  \textit{Ewen Camberlein$^1$, Diego di Carlo$^1$,}  \\
      \textit{Cl\'ement Gaultier$^1$, Antoine Deleforge$^2$, Nancy Bertin$^1$}}

\address{$^1$Univ Rennes 1, Inria, CNRS, IRISA, F-35000 Rennes, France \\
	$^2$Universit\'e de Lorraine, CNRS, Inria, Loria, F-54000 Nancy, France}

\begin{document}

\ninept
\maketitle

\begin{sloppy}

\begin{abstract}
This short paper presents an efficient, flexible implementation of the SRP-PHAT multichannel sound source localization method. The method is evaluated on the single-source tasks of the LOCATA 2018 development dataset, and an associated Matlab toolbox is made available online.
\end{abstract}


\section{Introduction}
\label{sec:intro}

Many source localization methods are based on the estimation of time-difference-of-arrival (TDOA) between microphones. In the two-microphone case, the thorough evaluation carried out in \cite{Blandin2012} highlighted the good performance of the Generalized Cross-Correlation with PHAse Transform (GCC-PHAT) algorithm \cite{knapp1976gcc} among several other methods. A general principle for extending this method to the multichannel setting, named SRP-PHAT, was proposed in \cite{Dibiase2001Robust}. However, no reference implementation of this family of algorithms (including all its possible variants) was available to date, to the best of the authors' knowledge. In particular, choices of parameters (pooling in time and frequency, grid search resolution) and management of the multiple coordinate systems
between pairs of microphones and the whole antenna
have not been systematically documented.


We present here our participation to the 2018 LOCATA challenge \cite{lollmann2018locata}. This aims at evaluating our implementation of SRP-PHAT, whose good performance was already probed in realistic environments in \cite{bertin2018voicehome2}, but which was not confronted to other methods on a common, realistic dataset so far. It is here applied to task 1 (single, static source and static array), task 3 (single moving source and static array) and task 5 (single moving source and moving array) of the challenge. Evaluation of the method on the development dataset shows encouraging results, confirming the legitimate place of SRP-PHAT among state-of-the-art methods for multichannel source localization in reverberant environments.

\section{Principles}
\label{sec:principles} 


\subsection{TDOA and angular spectrum}
The general principle of SRP-PHAT is to compute a function $\Phi(\theta,\varphi)$ termed ``angular spectrum'', where $\theta$ and $\phi$ are azimuth and elevation variables, which is expected to exhibit local maxima in the directions of active sources. More precisely, the following steps are followed:
\begin{itemize}
\item build in each time-frequency $(t,f)$ bin a local angular spectrum \emph{function} $\phi(t,f,\theta,\varphi)$ that is large for directions $(\theta,\varphi)$ which are compatible with the observed signal at $(t,f)$ and small otherwise,
\item \emph{integrate} (or \textit{pool}) this function over the time-frequency plane, leading to a global angular spectrum,
\item find the \emph{peaks} of this spectrum above a certain \emph{threshold} and distant from a certain \emph{minimum angle}.\\[1em]
\end{itemize}


\subsection{GCC-PHAT} 
The GCC algorithm \cite{knapp1976gcc} generalizes cross-correlation by multiplying the cross-spectral density (CSD) of two signals by complex weights in the frequency domain. The PHAT weighting consists of normalizing the CSD to have unit amplitude at all frequencies, thus considering phase differences only and minimizing the influence of the source's power spectral density. In the case of a single broadband source, two microphones and under far-field (source-array distance much larger that the array aperture $h$), free-field (only the direct source-to-microphone sound paths are considered) and noiseless assumptions, the asymptotic GCC-PHAT of the microphone signals in the discrete frequency domain is $e^{-2i\pi f \tau\cos(\alpha)}$ where $f$ is the frequency index, $\alpha$ is the angle of arrival (AOA) of the source signal at the array and $\tau$ is the maximum observable delay in samples. We have $\tau=hF_s/C$ where $C$ denotes the speed of sound and $F_s$ is the frequency of sampling. This leads to the following natural local angular spectrum for a microphone pair:

\begin{equation}
\label{eq:GCCPHAT}
\phi^{\rm GCC-PHAT}(t,f,\alpha) = \Re \left( \frac{x_1(t,f) x_2^*(t,f)}{|x_1(t,f) x_2^*(t,f)|} e^{-2i\pi f \tau\cos(\alpha)} \right)
\end{equation}
where $x_1$, $x_2$ denote the microphone signals in the short-time Fourier transform (STFT) domain and $\alpha$ is a \textit{local} azimuth in the coordinate system defined by the microphone pair. Note that this spectrum does not depend on elevation.

\subsection{Extension to multichannel}
\label{sec:srpphat}

SRP-PHAT's principle is to first compute local angular spectra for each microphone pair, then to aggregate them across pairs (bringing them back in the global coordinate system) before being pooled and maximized. The computation consists of the following steps:
\begin{enumerate}
\item Define the search space, \textit{i.e.} a grid of possible DOAs $(\theta_j,\varphi_k)$ for which we want to evaluate $\Phi$ in the global coordinate system; 
\item For each microphone pair $n$:
\begin{enumerate}
\item Compute the corresponding AOAs $\{\alpha_{jk}^{(n)}\}_{jk}$ with respect to the microphone pair;
\item Resample $\{\alpha_{jk}^{(n)}\}_{jk}$ into a smaller set $\{\alpha_i^{(n)}\}_i$ in order to reduce computational time;
\item Compute the GCC-PHAT local angular spectrum $\phi_n$ at angles $\{\alpha_i^{(n)}\}_i$ for microphone pair $n$ according to formula \eqref{eq:GCCPHAT};
\item Linearly interpolate $\phi_n$ back to the original angle resolution and global coordinate system;
\end{enumerate}
\item Compute the global spectrum $\Phi(\theta_j,\varphi_k)$ by pooling the local angular spectra $\phi_n$ over all time-frequency bins $(t,f)$ and across all microphone pairs $n$. Pooling methods (such as maximum or sum) and their order over each of the $f,t,n$ indexes must be chosen for this purpose.
\item Find the indexes $j$ and $k$ of the largest peak (single source case) or peaks (multiple source case) of $\Phi(\theta_j,\varphi_k)$, yielding the estimated source azimuth(s) $\theta_j$ and elevation(s) $\varphi_k$.
\end{enumerate}

\begin{table*}[!b]
	\centering
	\caption{Sound localization results obtained on the LOCATA development dataset using the proposed SRP-PHAT method. For each task and array, different \textit{success thresholds} (in degrees) are considered. A localization is considered successful when both azimuth and elevation estimation errors are below these thresholds. The percentage of successful localizations (suc.) and their average azimuth (az.) and elevation (el.) errors in degrees are showed for each threshold.}
	\begin{tabular}{|c|c|c|c|c|c|c|}
		\hline
		\multirow{3}{12mm}{Success threshold} & \multicolumn{2}{|c}{Task 1} & \multicolumn{2}{|c}{Task 3} & \multicolumn{2}{|c|}{Task 5} \\
		\cline{2-7}
		& Robot head & Eigenmike & Robot head & Eigenmike & Robot head & Eigenmike \\
		\cline{2-7}
		& \hspace{1mm}az.\hspace{1.6mm}$|$\hspace{1.3mm}el.\hspace{1.5mm}$|$\hspace{1.2mm}suc.
		& \hspace{1mm}az.\hspace{1.6mm}$|$\hspace{1.3mm}el.\hspace{1.5mm}$|$\hspace{1.2mm}suc.
		& \hspace{1mm}az.\hspace{1.6mm}$|$\hspace{1.3mm}el.\hspace{1.5mm}$|$\hspace{1.2mm}suc.
		& \hspace{1mm}az.\hspace{1.6mm}$|$\hspace{1.3mm}el.\hspace{1.5mm}$|$\hspace{1.2mm}suc.
		& \hspace{1mm}az.\hspace{1.6mm}$|$\hspace{1.3mm}el.\hspace{1.5mm}$|$\hspace{1.2mm}suc.
		& \hspace{1mm}az.\hspace{1.6mm}$|$\hspace{1.3mm}el.\hspace{1.5mm}$|$\hspace{1.2mm}suc. \\
		\hline
		No thresh. & $1.51|1.71|\hspace{1.6mm}-\hspace{1.5mm} $ & $7.04|4.68|\hspace{1.6mm}-\hspace{1.5mm} $ & $4.43|2.66|\hspace{1.6mm}-\hspace{1.5mm} $ & $8.79|4.41|\hspace{1.6mm}-\hspace{1.5mm} $ & $6.19|3.16|\hspace{1.6mm}-\hspace{1.5mm} $ & $9.31|4.37|\hspace{1.6mm}-\hspace{1.5mm} $ \\
		$20^{\circ}$ & $1.43|1.66|99.9$ & $6.95|4.64|99.9$ & $2.48|1.75|95.8$ & $7.82|3.12|92.5$ & $1.76|1.83|94.5$ & $5.84|2.94|94.8$ \\
		$10^{\circ}$ & $1.43|1.66|99.9$ & $6.95|4.64|99.9$ & $2.35|1.56|93.7$ & $6.27|2.40|71.0$ & $1.64|1.77|93.3$ & $5.43|2.80|89.5$ \\
		\hline	
	\end{tabular}
	\label{tab:dregon}
	\vspace{-6mm}
\end{table*}

%



A Matlab toolbox allowing easy and flexible implementations of SRP-PHAT as well as 7 other angular spectrum-based was made freely available online under the name Multichannel BSS Locate\footnote{\url{http://bass-db.gforge.inria.fr/bss_locate/}}.

\section{Experiments and Results}
\label{sec:results}
We now evaluate our implementation of SRP-PHAT on tasks 1, 3 and 5 of the LOCATA challenge \cite{lollmann2018locata} with the \textit{robot head} and the \textit{Eigenmike} arrays for azimuth and elevation estimation. The other two antennas of the challenge are discarded since they are near-linear (preventing elevation estimation). Moreover, DICIT is incompatible with our far-field assumption while the dummy head is incompatible with our free-field assumption. All microphones (12) and all microphone pairs (66) are used with the robot head. All microphones are also used with the Eigenmike, but microphone pairs with a curvilinear distance lesser than $90^{\circ}$ are discarded in order to reduce the overall algorithm complexity, resulting in 240 microphone pairs usage instead of the 496 available pairs. A sphere sampled with $1^{\circ}$ resolution both in azimuth and elevation is used as the search space (see Sec. \ref{sec:srpphat} - 1).
For each microphone pair, the evaluated AOAs (see Sec. \ref{sec:srpphat} - 2.(b)) are computed with $5^{\circ}$ resolution. SRP-PHAT is applied every 256 ms to 512 ms signals (downsampled to 16 kHz) using an overlapping sliding analysis window. The STFT is applied to each channel of the signals using 64 ms Fourier frames (1024 samples) with 50\% overlap and sine windows. This results in 15 Fourier frames per analysis window. The pooling methods used are first summations over microphone pairs and frequencies and then maximum over time frames. Only the largest value over the global angular search space is returned as a DOA estimate, as only single source localization is considered. DOA estimates are linearly interpolated over time in order to match the one-estimate-per-time-stamp requirement of LOCATA.

Table 1. shows localization errors obtained with the proposed method. Encouragingly, mean azimuth errors (first row) are 2 to 6 times smaller than those reported for the baseline MUSIC method \cite{lollmann2018locata} for the same arrays and tasks. To reduce the effect of gross errors on reported mean values, average \textit{successful localization} errors for different success threshold are also reported. The proposed method localizes the target source with less than $20^\circ$ error in at least $92\%$ of the tests for all tasks and arrays. Nearly 100\% correct localization is achieved in the static scenario (task 1). While the robot head array enables better performance in all tasks, average errors never exceed $10^{\circ}$ regardless of the task and array.

\section{Conclusion}
\label{sec:conclusion}
In this paper, we reported competitive single sound source localization results using the SRP-PHAT implementation of the Multichannel BSS Locate toolbox on the 2018 LOCATA challenge dataset. The toolbox's flexibility allows one to easily adapt the method to arbitrary array geometries, to various types of emitted signals and environments, and to easily generalize any two-channel localization method to more channels. It can also be tuned to find a trade-off between computational time and accuracy, although we focused on accuracy for this challenge. While the toolbox can output multiple source direction estimates for a given input signal, it does not yet incorporate source counting or source tracking solutions, preventing its use in multiple sources scenario. Such extensions will be considered for future releases of the toolbox.
\bibliographystyle{IEEEtran}
\footnotesize
\bibliography{Locata2018}
%
%
%
%
%
%
%
%
%

\end{sloppy}
\end{document}